\documentclass[10pt,aps,twocolumn,floatfix,showpacs,superscriptaddress]{revtex4-1}
\usepackage{graphicx,amsmath,amsfonts,mathrsfs}
\usepackage{bm}
\usepackage[]{color}

\begin{document}
\title{Skyrmionic order and magnetically induced polarization change in lacunar spinel compounds GaV$_{4}$S$_{8}$ and GaMo$_{4}$S$_{8}$: comparative theoretical study}
\author{S.~A.~Nikolaev}
\email{saishi@inbox.ru}
\affiliation{Institute of Innovative Research, Tokyo Institute of Technology, 4259 Nagatsuta, Midori, Yokohama 226-8503, Japan}
\affiliation{National Institute for Materials Science, MANA,
1-1 Namiki, Tsukuba, Ibaraki 305-0044, Japan}
\affiliation{Department of Theoretical Physics and Applied Mathematics, Ural Federal University,
Mira str. 19, 620002 Ekaterinburg, Russia}
\author{I.~V.~Solovyev}
\email{SOLOVYEV.Igor@nims.go.jp}
\affiliation{National Institute for Materials Science, MANA,
1-1 Namiki, Tsukuba, Ibaraki 305-0044, Japan}
\affiliation{Department of Theoretical Physics and Applied Mathematics,
Ural Federal University, Mira str. 19, 620002 Ekaterinburg, Russia}
\affiliation{Institute of Metal Physics, S. Kovalevskaya str. 18, 620108 Ekaterinburg, Russia}
\date{\today}

\begin{abstract}
We show how low-energy electronic models derived from the first-principles electronic structure calculations can help to rationalize the magnetic properties of two lacunar spinel compounds Ga$M_{4}$S$_{8}$ with relatively light ($M=$ V) and heavy ($M=$ Mo) transition-metal elements, which are responsible for different spin-orbit interaction strength. In the model, each magnetic lattice point was associated with the ($M_4$S$_4$)$^{5+}$ molecule, and the model itself was formulated in the basis of \emph{molecular} Wannier functions constructed for three magnetic $t_{2}$ bands. The effects of rhombohedral distortion, spin-orbit interaction, band filling, and the screening of Coulomb interactions in the $t_{2}$ bands are discussed in details by stressing similarities and differences between GaV$_{4}$S$_{8}$ and GaMo$_{4}$S$_{8}$. The electronic model is further treated in the superexchange approximation, which allows us to derive an effective spin model for the energy \emph{and} electric polarization ($\boldsymbol{P}$) depending on the relative orientation of spins in the bonds, and study the properties of this model by means of classical Monte Carlo simulations with the emphasis on the possible formation of the skyrmionic phase. While isotropic exchange interactions clearly dominate in GaV$_{4}$S$_{8}$, all types of interactions -- isotropic, antisymmetric, and symmetric anisotropic -- are comparable in the case of GaMo$_{4}$S$_{8}$. Particularly, large uniaxial exchange anisotropy has a profound effect on the properties of GaMo$_{4}$S$_{8}$. On the one hand, it raises the Curie temperature by opening a gap in the spectrum of magnon excitations. On the other hand, it strongly affects the skyrmionic phase by playing the role of a molecular field, which facilitates the formation of skyrmions, but makes them relatively insensitive to the external magnetic field in the large part of the phase diagram. We predict reversal of the magnetic dependence of $\boldsymbol{P}$ in the case of GaMo$_{4}$S$_{8}$ caused by the reversal of direction of the rhombohedral distortion.
\end{abstract}
\maketitle

\section{\label{sec:Intro} Introduction}
\par Magnetic skyrmions -- the topologically protected whirl-like spin textures -- have attracted great deal of attention~\cite{bog1,bog2,NagaosaTokura}. Owing to their topology and nanometer size, skyrmions behave like particle objects that can be moved over macroscopic distances by applying low-density electric currents~\cite{sk2,sk3}, thus making them suitable candidates for applications in low-power nanoelectronics and data storage~\cite{sk4}. Moreover, the studies of novel skyrmionic phases present a fundamental interest as they open new frontiers in our understanding of magnetic matter.

\par The existence of skyrmions has been theoretically predicted to occur in solids belonging to certain crystallographic classes, which allow for chiral magnetic structures driven by antisymmetric Dzyaloshinskii-Moriya (DM) interactions~\cite{bog1}. The skyrmions can be of two types: (i) Bloch skyrmions, where spins are locked in a tangential plane (and rotate in this plane), and (ii) N\'eel skyrmions with the spins rotating in radial planes.

\par The Bloch skyrmions are typically observed in metallic (and, therefore, non-polar) alloys including MnSi~\cite{sk1}, Fe$_{1-x}$Co$_{x}$Si~\cite{sk5}, and FeGe~\cite{sk3}. The N\'eel skyrmions were reported only recently in two materials with the lacunar spinel structure ($R3m$, the space group No. 160): GaV$_4$S$_8$~\cite{gavs1,gavs2} and GaV$_4$Se$_8$~\cite{gavse}. The new aspect of the lacunar spinels structure is that it is \emph{polar} and, therefore, the compounds are multiferroics. The multiferroicity adds a new functionality into the properties of skyrmions, including an electric-field control of these magnetic objects and inversely -- the control of electric polarization by changing the magnetic texture. The possibility of such control was indeed demonstrated by Ruff \textit{et al.}~\cite{gavs2}, who have shonw that the change of electric polarization in GaV$_4$S$_8$, caused by the change of the magnetic order, can reach several tens of $\mu$C/m$^2$. The only material where the multiferroicity was known to coexist with the skyrimon order was Cu$_{2}$OSeO$_{3}$~\cite{cuoseo1,cuoseo4,cuoseo3}. However, the observed magnetoelectric coupling was almost two orders of magnitude weaker than in GaV$_4$S$_8$.

\par Despite genuine interest in multiferroic skyrmions, the understanding of this phenomenon is still in a preliminary stage. It is not clear why the polarization depends on the skyrmionic texture, which microscopic mechanism is responsible for the polarization change, and how this property can be further controlled and improved.

\par In the previous communication~\cite{PRB2019}, we reported results of our first theoretically study of the electric polarization ($\boldsymbol{P}$) in GaV$_4$S$_8$ depending on the change of the skyrmion order. For these purposes, we started with the first-principles electronic structure calculations and established a realistic model describing the behavior of the magnetic $t_{2}$ bands near the Fermi level in the basis of appropriate molecular-type Wannier orbitals. In order to solve this model, we have extended the superexchange (SE) theory~\cite{Anderson} to deal not only with the exchange interactions but also with the change of electric polarization depending on the relative direction of spins in the bonds. Thus, this theory allowed us to construct a spin model for both the energy and $\boldsymbol{P}$, and then to study this model by using various techniques. By doing this, we were able to rationalize the behavior of electric polarization in GaV$_4$S$_8$. Particularly, (i) although the magnetic skyrmions are mainly formed by the SE interactions in the plane, another important factor, which determines the dependence of electric polarization on the magnetic order, is the stacking of these planes in the perpendicular direction $z$. In the lacunar spinel structure, the stacking is such that some neighboring spins in the adjacent planes remain noncollinear and this noncollinearity participates as the main source of the magnetic dependence of $\boldsymbol{P}$. (ii) Similar to the spin Hamiltonian, the magnetic part of the polarization can be decomposed in terms of isotropic, antisymmetric, and symmetric anisotropic contributions. In the case of N\'eel skyrmions, the magnetic dependence of $\boldsymbol{P}$ stems from the strong competition of the former two effects, emerging in 2nd order of spin-orbit (SO) coupling.

\par In the present article, we explain the details of our method. Furthermore, we extend our analysis to new lacunar spinel compound with strong SO coupling, GaMo$_{4}$S$_{8}$, which can potentially host the skyrmionic states~\cite{GaMo4S8Picozzi,Kitchaev,MGarst}. We will argue that the new aspect of GaMo$_{4}$S$_{8}$ is the strong exchange anisotropy, which favors the out-of-plane direction of spins and thus acts as a molecular field stabilizing the N\'eel skyrmions, but making them relatively insensitive to the external field in the large part of the phase diagram. In fact, all exchange interactions - isotropic, antisymmetric DM, and symmetric anisotropic - are comparable in the case of GaMo$_{4}$S$_{8}$, thus excluding any perturbative treatment. In comparison with GaV$_{4}$S$_{8}$, we predict the reversal of magnetic dependence of $\boldsymbol{P}$ in GaMo$_{4}$S$_{8}$, associated with reversal of the rhombohedral distortion.

\par The rest of the article is organized as follows. In Sec.~\ref{sec:LDA}, we briefly explain details of the crystal structure and basic electronic structure of GaV$_{4}$S$_{8}$ and GaMo$_{4}$S$_{8}$ within local density approximation (LDA)~\cite{lda}. In Sec.~\ref{sec:emodel}, we discuss construction and parameters of the electronic model for the molecular $t_{2}$ bands near the Fermi level. In Sec.~\ref{sec:smodel}, we present the spin model derived in the SE approximation for the magnetic interactions \emph{and} electric polarization. Results of Monte Carlo (MC) simulations for the spin model are discussed in Sec.~\ref{sec:Phase} and the brief summary of our work is given in Sec.~\ref{sec:summary}. Two appendices (\ref{sec:MC} and \ref{sec:HF}) deal with details of MC calculations and alternative estimates of parameters of the spin model based on the direct solution of the electronic model in the Hartree-Fock approximation.

\section{\label{sec:LDA} Crystal and basic electronic structure}
\par
\begin{figure}[b]
\begin{center}
\includegraphics[width=0.48\textwidth]{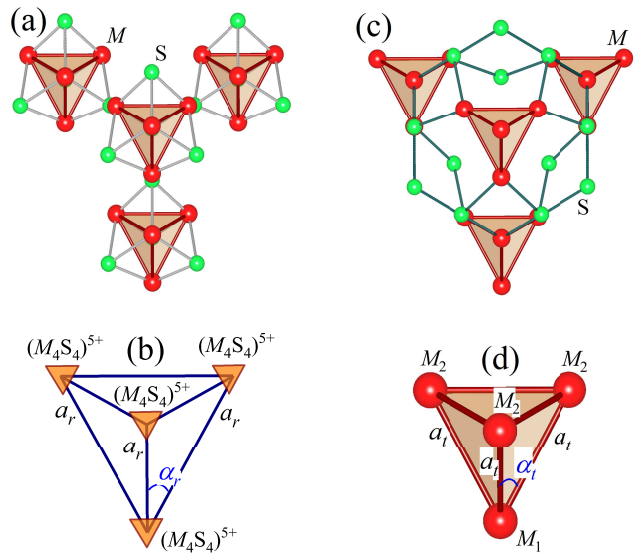}
\end{center}
\caption{Fragments of the crystal structure of Ga$M_4$S$_8$: (a) Network of the $M_4$S$_4$ ``molecules''; (b) Schematic view on the network with the notation of rhombohedral parameters $a_{r}$ and $\alpha_{r}$; (c) $M_4$ tetrahedra interconnected by S atoms; (d) Single $M_4$ tetrahedron with the notation of two inequivalent types of $M$ atoms, the $M_1$-$M_2$ bond length $a_{t}$ and the angle $\alpha_{t}$, characterizing the distortion of this tetrahedron.}
\label{fig.str}
\end{figure}

 The building blocks of the magnetic lattice of Ga$M_4$S$_8$ ($M=$ V and Mo) are charged ($M_4$S$_4$)$^{5+}$ ``molecules'', which are formed by two interpenetrating $M_4$ and S$_4$ tetrahedra. The molecules form the face-centered cubic network, as shown in Fig.~\ref{fig.str}(a) and (b), and are interconnected via yet another type of S atoms, as shown in Fig.~\ref{fig.str}(c). Below $T_{\rm s}$ (of about $44$ and $45$ K for GaV$_4$S$_8$ and GaMo$_4$S$_8$, respectively~\cite{GVS_struc,GMS_struc}) the lacunar spinels undergo a phase transition from the cubic $F\overline{4}3m$ to polar rhombohedral $R3m$ structure, which results in the deviation of the rhombohedral angle $\alpha_{r}$ from the ideal cubic value of $60^{\circ}$. Similar change is found for the angle $\alpha_{t}$, characterizing the distortion of the single $M_4$ tetrahedron, as explained in Fig.~\ref{fig.str}(d).
\noindent

\par In this study we use experimental parameters of the crystal structure for GaV$_4$S$_8$ and GaMo$_4$S$_8$ reported in Refs.~\cite{GVS_struc} and \cite{GMS_struc}, respectively. Unless otherwise stated, we focus on the behavior of the low-temperature $R3m$ phases. Some of parameters of the $R3m$ structures are listed in Table~\ref{tab:structure}.
\noindent
\begin{table}[h!]
\caption{Crystal-structure parameters of Ga$M_4$S$_8$ in their low-temperature $R3m$ phases (see Fig.~\ref{fig.str}): rhombohedral lattice parameter $a_{r}$ (in \AA), rhombohedral angle $\alpha_{r}$ (in $^{\circ}$), and the unit cell volume $V$ (in \AA$^3$). The parameters of the single $M_4$ tetrahedron (the $M_1$-$M_2$ distance, $a_{t}$, the $M_2$-$M_1$-$M_2$ angle, $\alpha_{t}$, and the volume, $V_{t}$) are given for comparison in parentheses.}
\label{tab:structure}
\begin{ruledtabular}
\begin{tabular}{lccc}
              & $a_{r}$ ($a_{t}$)      & $\alpha_{r}$ ($\alpha_{t}$) & $V$ ($V_{t}$) \\
\hline
GaV$_4$S$_8$  & $6.834$ ($2.898$)  &    $59.66$ ($58.36$)            & $223.95$ ($2.76$)     \\
GaMo$_4$S$_8$ & $6.851$ ($2.823$)  &    $60.53$ ($61.51$)            & $230.08$ ($2.74$)
\end{tabular}
\end{ruledtabular}
\end{table}
\noindent Particularly, we note that the direction of the rhombohedral distortion is opposite in the V- and Mo-based compounds: while GaV$_4$S$_8$ is elongated along the cubic $[111]$ axis ($\alpha_{r} < 60^{\circ}$), GaMo$_4$S$_8$ is compressed ($\alpha_{r} > 60^{\circ}$). Similar tendency is seen for the single $M_4$ tetrahedron. The unit cell volume is substantially larger in GaMo$_4$S$_8$ even though the single Mo$_4$ tetrahedron is smaller than V$_4$. Therefore, the Mo$_4$ octahedra are more compressed (thus, resulting in larger molecular level-splitting), but more separated from each other in comparison with V$_4$ in GaV$_4$S$_8$. In the following, we will show that all these changes are clearly reflected in the electronic structure and parameters of spin models of the considered lacunar spinel compounds.

\par
\begin{figure}[b]
\begin{center}
\includegraphics[width=0.48\textwidth]{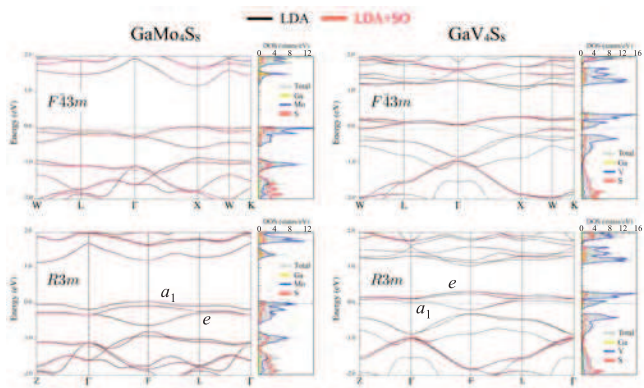}
\end{center}
\caption{Electronic structure and densities of states of GaMo$_4$S$_8$ and GaV$_4$S$_8$ in the local density approximation as obtained for the high-temperature cubic $F\overline{4}3m$ phase and the low-temperature rhombohedral $R3m$ phase with and without spin-orbit interaction. The Fermi level is at zero energy (shown by dashed line).}
\label{fig.LDA}
\end{figure}

 The electronic band structure of GaV$_4$S$_8$ and GaMo$_4$S$_8$, calculated within LDA using \texttt{Quantum ESPRESSO} method~\cite{qe}, is summarized in Fig.~\ref{fig.LDA}.
\noindent
\noindent Despite complexity of the lacunar systems, the electronic structure near the Fermi level is relatively simple and featured by three $t_{2}$ bands, which are separated by energy gaps from other bands located in the lower and upper energy parts of the spectrum. Importantly, these bands have a ``molecular origin'', resulting from the following hierarchy of hybridization effects. (i) The strong hybridization within the ($M_4$S$_4$)$^{5+}$ molecules leads to the formation of the molecular levels. (ii) The hybridization between the molecules is considerably weaker, resulting in the formation of weakly dispersive bands within each group of the molecular orbitals (but not in the overlap between the bands formed by different groups of the molecular orbitals). In the following, we will call the $t_{2}$ bands as ``target bands'', implying that the magnetism of GaV$_4$S$_8$ and GaMo$_4$S$_8$ originates mainly from this group of states and can be described by a proper model with all the parameters extracted from first-principles electronic structure calculations. Without SO interaction, the $t_{2}$ bands in the high-temperature cubic $F\overline{4}3m$ phase are threefold degenerate, while the rhombohedral distortion in the $R3m$ phase lifts this degeneracy and splits the $t_{2}$ bands into the singly degenerate $a_{1}$ and twofold degenerate $e$ bands. The splitting is clearly seen in Fig.~\ref{fig.LDA}. The $a_{1}$ band lies below the $e$ bands in GaV$_4$S$_8$ and above them in GaMo$_4$S$_8$, thus reflecting the change of the direction of the rhombohedral distortion. Taking into account the formal occupancy of the $t_{2}$ states, having one electron in GaV$_4$S$_8$ and one hole in GaMo$_4$S$_8$, the band splitting is consistent with general Jahn-Teller theorem saying that the rhombohedral distortion should lift the degeneracy of the ground state. Judging from the band dispersion alone, the SO interaction does not seem to play a decisive role: the change of the band structure caused by the SO interaction is relatively small compared to the effect of the rhombohedral distortion, even in GaMo$_4$S$_8$. Nevertheless, this interaction is vitally important for the formation of the skyrmion texture as it gives rise to such key ingredients as DM interactions and uniaxial anisotropy.

\section{\label{sec:emodel} Effective electronic model}
\par Our next goal is the construction of realistic model, which would describe the magnetic properties of GaV$_4$S$_8$ and GaMo$_4$S$_8$. Since the skyrmionic texture can include thousands of atoms in the magnetic unit cell, the brute-force electronic structure calculations, dealing with such complex noncollinear magnetic states, are practically impossible today. Nevertheless, one can construct a model, using the input from the electronic structure calculations, and then solve this model. Another problem is the electronic correlations in the molecular complexes ($M_4$S$_4$)$^{5+}$, which can be relatively easily taken into account in the model, but not at the level of first-principles electronic structure calculations.

\par In order to fulfil this goal, we first construct the basis of Wannier functions for the $t_{2}$ bands, using the maximally localized Wannier functions technique~\cite{WannierRevModPhys} as implemented in the \texttt{wannier90} package~\cite{wan90}. Thus, in our case, the Wannier functions are the molecular orbitals of the ($M_4$S$_4$)$^{5+}$ clusters of the $a_{1}$ and $e$ symmetry. Table~\ref{tab:wannier} summarizes the spreads of these Wannier functions, which characterize the degree of their localization~\cite{WannierRevModPhys}.
\noindent
\begin{table}[t]
\caption{Spreads of the Wannier functions (in \AA$^2$) corresponding to $a_{1}$ and $e$ representations.}
\label{tab:wannier}
\begin{ruledtabular}
\begin{tabular}{lcc}
              & $a_{1}$       &   $e$       \\
\hline
GaV$_4$S$_8$  & $8.4$         &   $8.1$     \\
GaMo$_4$S$_8$ & $9.4$         &   $9.7$
\end{tabular}
\end{ruledtabular}
\end{table}
\noindent The Wannier functions for GaMo$_4$S$_8$ tend to be more extended despite a smaller Mo$_4$ tetrahedron size. Nevertheless, this can be easily explained by the character of atomic $4d$ orbitals, which are less localized in comparison with the $3d$ ones. Then, the direction of the rhombohedral distortion also affects the relative localization of the $a_{1}$ and $e$ orbitals: while the $a_{1}$ orbital is the least localized in GaV$_4$S$_8$, it becomes the most localized in GaMo$_4$S$_8$.

\par Then, the Wannier functions are used as the basis for the construction of the low-energy model~\cite{JPCMreview,PRB2019}:
\noindent
\begin{equation}
\hat{\mathcal{H}}^{\mathrm{el}}=\hat{\mathcal{H}}_{\mathrm{CF}}+\hat{\mathcal{H}}_{\mathrm{SO}}+\hat{\mathcal{H}}_{\mathrm{kin}}+\hat{\mathcal{H}}_{U},
\label{eq:elmodel}
\end{equation}
\noindent
where the first three terms (the crystal field, the spin-orbit interaction, and the kinetic hoppings, respectively) is the noninteracting one-electron part of the model Hamiltonian and $\hat{\mathcal{H}}_{U}$ stands for the effective electron-electron interactions in the $t_{2}$ band. In our model, the one-electron part was defined via matrix elements of the LDA Hamiltonian in the Wannier basis~\cite{JPCMreview}, while the electron-electron interaction part was evaluated within the constrained random phase approximation (cRPA)~\cite{rpa2}.

\par Thus, $\hat{\mathcal{H}}^{\mathrm{el}}$ is formulated in terms of creation (annihilation) operators $\hat{c}_{ia}^{\sigma\dagger}$ ($\hat{c}_{ia}^{\sigma\phantom{\dagger}}$) of an electron with the spin $\sigma$ at the molecular Wannier orbital $a$ of the site $i$ (where $a=1$ is the $a_{1}$ orbital, while $a=2$ and $3$ form the basis of the two-dimensional representation $e$). Particularly, we define the crystal-field splitting as $\hat{\mathcal{H}}_{\mathrm{CF}}=\Delta\,\sum_{i,a \ne 1,\sigma} \hat{c}_{ia}^{\sigma\dagger}\hat{c}_{ia}^{\sigma\phantom{\dagger}}$ and the SO interaction as $\hat{\mathcal{H}}_{\mathrm{SO}} = \zeta_{SO} \sum_{i} \hat{\boldsymbol{L}}_{i} \cdot \hat{\boldsymbol{S}}_{i} - \zeta_{SO}^{R} \sum_{i} \left( \hat{L}^{x}_{i}\hat{S}^{x}_{i} + \hat{L}^{y}_{i}\hat{S}^{y}_{i} \right)$~\cite{PRB2019}, where the first term stands for the regular ``spherical'' interaction while the second term is the Rashba-type interaction arising from the polar rhombohedral distortion~\cite{rashba}. The matrix elements of angular momentum operators are expressed in terms of the antisymmetric Levi-Civita symbol as $( \hat{L}^{x}_{i} )^{ab} = -i\varepsilon_{2ab}$, $( \hat{L}^{y}_{i} )^{ab} = -i\varepsilon_{3ab}$, and $( \hat{L}^{z}_{i} )^{ab} = i\varepsilon_{1ab}$. The corresponding parameters are listed in Table~\ref{tab:onsite}.
\noindent
\begin{table}[b]
\caption{Parameters of crystal-field splitting, $\Delta$, and spin-orbit interaction of the spherical type, $\zeta_{SO}$, and Rashba type $\zeta_{SO}^{R}$ (all are in meV).}
\label{tab:onsite}
\begin{ruledtabular}
\begin{tabular}{lccc}
              & $\Delta$            &   $\zeta_{SO}$    &   $\zeta_{SO}^{R}$    \\
\hline
GaV$_4$S$_8$  & $\phantom{-1}98.1$  &   $23.0$          &   $\phantom{-}1.3$    \\
GaMo$_4$S$_8$ & $-168.0$            &   $68.7$          &   $-8.7$
\end{tabular}
\end{ruledtabular}
\end{table}
\noindent First, we note that the sign of $\Delta$ and $\zeta_{SO}^{R}$ is controlled by the direction of the rhombohedral distortion: both parameters are positive in GaV$_4$S$_8$, where $\alpha_{r} < 60^\circ$, but become negative in GaMo$_4$S$_8$, where $\alpha_{r} > 60^\circ$. The molecular level-splitting is due to the hybridization effects within each ($M_4$S$_4$)$^{5+}$ cluster~\cite{kanamori}, which are stronger in GaMo$_4$S$_8$ because (i) the Mo$_4$ tetrahedron is smaller and (ii) the Mo $4d$ states are more extended, which explain larger value of $| \Delta |$. The SO coupling $\zeta_{SO}$ is also larger in GaMo$_4$S$_8$, as expected for heavier Mo atoms. The large value of $| \zeta_{SO}^{R} |$ in GaMo$_4$S$_8$ is a joint effect of hybridization and relativistic interactions associated with the Mo states.

\par
\begin{figure}[b]
\begin{center}
\includegraphics[width=0.48\textwidth]{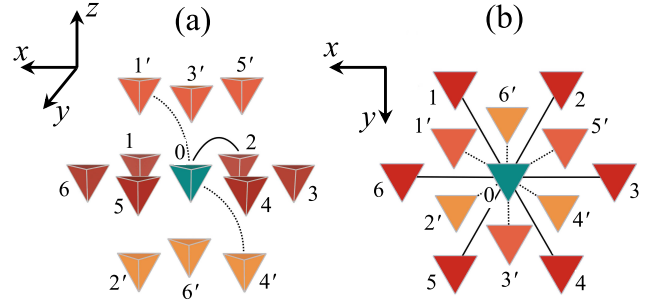}
\end{center}
\caption{Notations of $M_4$ clusters in Ga$M_4$S$_8$: (a) side view, (b) top view.}
\label{fig.notations}
\end{figure}

 The kinetic part, $\hat{\mathcal{H}}_{\mathrm{kin}}=\sum_{i \ne j} \sum_{ab\sigma} t_{ij}^{ab}\hat{c}_{ia}^{\sigma\dagger}\hat{c}_{jb}^{\sigma\phantom{\dagger}}$, is given by the transfer integrals $\hat{t}_{ij} = [t_{ij}^{ab}]$. For the in-plane bonds ($j=1$-$6$ in Fig.~\ref{fig.notations}), they can be presented as
\noindent
\noindent
\begin{widetext}
\begin{equation}
\hat{t}_{0j} =
\left(
\begin{array}{ccc}
\mathit{t}_{\parallel}^{1} & \mathit{s}_{\parallel}^{3} \sin \frac{2\pi j}{3} - \mathit{u}_{\parallel}^{3} \cos \frac{\pi j}{3} & -\mathit{s}_{\parallel}^{3} \cos \frac{2\pi j}{3} + \mathit{u}_{\parallel}^{3} \sin \frac{\pi j}{3} \\
\mathit{s}_{\parallel}^{3} \sin \frac{2\pi j}{3} + \mathit{u}_{\parallel}^{3} \cos \frac{\pi j}{3} & \mathit{t}_{\parallel}^{2} - \mathit{s}_{\parallel}^{2} \cos \frac{2\pi j}{3} & \mathit{s}_{\parallel}^{2} \sin \frac{2\pi j}{3} + (-1)^{j} \mathit{u}_{\parallel}^{2} \\
-\mathit{s}_{\parallel}^{3} \cos \frac{2\pi j}{3} - \mathit{u}_{\parallel}^{3} \sin \frac{\pi j}{3} & \mathit{s}_{\parallel}^{2} \sin \frac{2\pi j}{3} - (-1)^{j} \mathit{u}_{\parallel}^{2} & \mathit{t}_{\parallel}^{2} + \mathit{s}_{\parallel}^{2} \cos \frac{2\pi j}{3}
\end{array}
\right).
\label{eq:tin}
\end{equation}
\end{widetext}
\noindent Six independent parameters, describing (i) the diagonal bond-independent hoppings between orbitals of either $a_{1}$ or $e$ symmetry ($\mathit{t}_{\parallel}^{1}$ and $\mathit{t}_{\parallel}^{2}$, respectively); (ii) the symmetric ($\mathit{s}_{\parallel}^{2}$) and antisymmetric ($\mathit{u}_{\parallel}^{2}$) hoppings between different $e$ orbitals ; and (iii) the symmetric ($\mathit{s}_{\parallel}^{3}$) and antisymmetric ($\mathit{u}_{\parallel}^{3}$) hoppings connecting $a_{1}$ and one of $e$ orbitals are listed in Table~\ref{tab:tin}.
\noindent
\begin{table}[t]
\caption{Hopping parameters for the nearest-neighbor in-plane bonds (in meV). The corresponding $3$$\times$$3$ matrices of transfer integrals are given by Eq.~(\ref{eq:tin}).}
\label{tab:tin}
\begin{ruledtabular}
\begin{tabular}{lrrrrrr}
              & $\mathit{t}_{\parallel}^{1}$ & $\mathit{s}_{\parallel}^{3}$ & $\mathit{u}_{\parallel}^{3}$ & $\mathit{t}_{\parallel}^{2}$ & $\mathit{s}_{\parallel}^{2}$ & $\mathit{u}_{\parallel}^{2}$ \\
\hline
GaV$_4$S$_8$  & $4.0$                        &   $25.5$                     &              $16.2$          &           $-0.4$             &   $-10.5$                    &              $18.7$          \\
GaMo$_4$S$_8$ & $7.3$                        &   $37.4$                     &             $-14.8$          &            $0.3$             &   $-15.6$                    &             $-24.0$
\end{tabular}
\end{ruledtabular}
\end{table}

\par The matrices of transfer integrals for the out-of-plane bonds ($j=1'$-$6'$ in Fig.~\ref{fig.notations}) are described by five independent parameters $\mathit{t}_{\perp}^{1}$, $\mathit{t}_{\perp}^{2}$, $\mathit{s}_{\perp}^{2}$, $\mathit{s}_{\perp}^{3}$, and $\mathit{u}_{\perp}^{3}$, which have the same meaning as for the in-plane bonds~\cite{footnote3}:
\noindent
\begin{widetext}
\begin{equation}
\hat{t}_{0j} =
\left(
\begin{array}{ccc}
\mathit{t}_{\perp}^{1} & \mathit{s}_{\perp}^{3} \sin \frac{2\pi j}{3} - \mathit{u}_{\perp}^{3} \sin \frac{\pi j}{3} & \mathit{s}_{\perp}^{3} \cos \frac{2\pi j}{3} + \mathit{u}_{\perp}^{3} \cos \frac{\pi j}{3} \\
\mathit{s}_{\perp}^{3} \sin \frac{2\pi j}{3} + \mathit{u}_{\perp}^{3} \sin \frac{\pi j}{3} & \mathit{t}_{\perp}^{2} + \mathit{s}_{\perp}^{2} \cos \frac{2\pi j}{3} & \mathit{s}_{\perp}^{2} \sin \frac{2\pi j}{3} \\
\mathit{s}_{\perp}^{3} \cos \frac{2\pi j}{3} - \mathit{u}_{\perp}^{3} \cos \frac{\pi j}{3} & \mathit{s}_{\perp}^{2} \sin \frac{2\pi j}{3} & \mathit{t}_{\perp}^{2} - \mathit{s}_{\perp}^{2} \cos \frac{2\pi j}{3}
\end{array}
\right).
\label{eq:tout}
\end{equation}
\end{widetext}
\noindent These parameters are listed in Table~\ref{tab:tout}.
\noindent
\begin{table}[b]
\caption{Hopping parameters for the nearest-neighbor out-of-plane bonds (in meV). The corresponding $3$$\times$$3$ matrices of transfer integrals are given by Eq.~(\ref{eq:tout}).}
\label{tab:tout}
\begin{ruledtabular}
\begin{tabular}{lrrrrr}
              & $\mathit{t}_{\perp}^{1}$ & $\mathit{s}_{\perp}^{3}$ & $\mathit{u}_{\perp}^{3}$ & $\mathit{t}_{\perp}^{2}$ & $\mathit{s}_{\perp}^{2}$ \\
\hline
GaV$_4$S$_8$  & $-3.3$                   &   $-22.7$                &              $-21.6$     &           $2.3$          &   $21.7$                 \\
GaMo$_4$S$_8$ & $-4.7$                   &   $-25.3$                &              $25.5$      &           $5.7$          &   $29.9$
\end{tabular}
\end{ruledtabular}
\end{table}
\noindent Without SO interactions, the only relevant parameters are $\mathit{t}^{1}$, $\mathit{s}^{3}$, and $\mathit{u}^{3}$, which involve the occupied $a_{1}$ orbital. For instance, only these parameters will contribute to the exchange coupling and the electric polarization in the framework of the SE theory~\cite{PRB2019}. Quite expectedly, these transfer integrals are stronger in GaMo$_4$S$_8$, due to the spacial extension of the Mo $4d$ states. The antisymmetric part of $\hat{t}_{ij}$, described by $u^{2}$ and $u^{3}$, is an odd function of the rhombohedral distortion. Therefore, $u^{2}$ and $u^{3}$ have different signs in GaV$_4$S$_8$ and GaMo$_4$S$_8$, where this distortion has opposite directions.

\par
\begin{figure}[b]
\begin{center}
\includegraphics[width=0.48\textwidth]{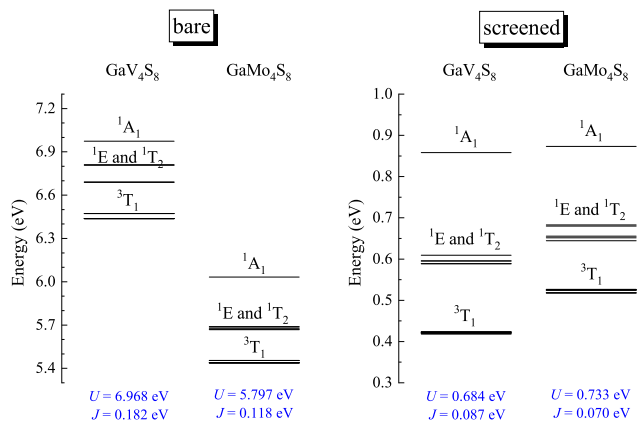}
\end{center}
\caption{Energies of two-particle excitations obtained using matrix elements of bare (left) and screened (right) Coulomb interactions and corresponding averaged Kanamori parameters of intraorbital Coulomb interaction $U$ and Hund's rule exchange interaction $J$.}
\label{fig.multiplet}
\end{figure}

 Finally, the electron-electron interaction in (\ref{eq:elmodel}) is given by
\begin{equation}
\hat{\mathcal{H}}_{U} = \frac{1}{2}
\sum_{i}  \sum_{\sigma \sigma'} \sum_{abcd} U^{abcd}
\hat{c}^{\sigma\dagger}_{i a } \hat{c}^{\sigma' \dagger}_{i c }
\hat{c}^{\sigma \phantom{\dagger}}_{i b }
\hat{c}^{ \sigma' \phantom{\dagger}}_{i d},
\label{eq:hublow}
\end{equation}
\noindent where the screened Coulomb interactions, $\hat{U} = [ U^{abcd} ]$, were calculated within cRPA~\cite{rpa2}. In Fig.~\ref{fig.multiplet}, we show the energies of two-particle excitations (i.e., ``two-electron'' in the case of GaV$_4$S$_8$ and ``two-hole'' in the case of GaMo$_4$S$_8$), obtained for bare and screened interactions (and for the time being ignoring the crystal field and spin-orbit interaction). These energies are relevant to the superexchange processes, which will be considered below.
\noindent
\noindent In the ideal spherical case, the two-particle states are split in three groups: nine $^3\mathrm{T}_1$ states, the degenerate $^1\mathrm{T}_2$ and $^1\mathrm{E}$ states (five in total), and one $^1\mathrm{A}_1$ state with the energies $(U$$-$$3J)$, $(U$$-$$J)$, and $(U$$+$$2J)$, respectively~\cite{Oles2005}, which are given in terms of the Kanamori's intraorbital Coulomb interaction $U$ and exchange interaction $J$~\cite{Kanamori}. The rhombohedral distortion and covalent mixing~\cite{Vaugier2012,Ribic2014}, manifested in different spreads of the $a_{1}$ and $e$ Wannier functions (Table~\ref{tab:wannier}), partially lift the degeneracy of the $^3\mathrm{T}_1$, $^1\mathrm{T}_2$, and $^1\mathrm{E}$ states. The averaged Kanamori parameters, evaluated using the energetic centers of gravity of these states, are also shown in Fig.~\ref{fig.multiplet}. First, we note that, due to the spacial extension of the molecular orbitals, the bare $U \sim$ $6$-$7$ eV is substantially smaller than typical atomic values of $U$ (of the order of $20$ eV). Furthermore, the behavior of bare interactions fully reflects the degree of the localization of the Wannier functions, where the bare $U$ decreases in the direction GaV$_4$S$_8$ $\to$ GaMo$_4$S$_8$, following the increase of the Wannier functions spreads (Table~\ref{tab:wannier}). Similar behavior is found for bare $J$, which decreases drastically in comparison with the atomic values (about $0.8$ eV for V and $0.5$ eV for Mo), but still appears to be larger in the case of more localized V $3d$ states in GaV$_4$S$_8$. Nevertheless, even more important is the screening, which substantially modifies the behavior of $U$ and $J$. Particularly, the screening of Coulomb interactions is exceptionally strong due to proximity of other occupied and unoccupied bands to the target $t_{2}$ bands (see Fig.~\ref{fig.LDA}). Moreover, all these bands are basically the transition-metal $d$ bands, which makes the screening very efficient~\cite{PRB2005}. As the result, the screened $U$ is reduced by an order of magnitude till about $0.7$ eV in both GaV$_4$S$_8$ and GaMo$_4$S$_8$. The same screening reduces $J$'s by a factor 2 compared to their bare values. Although the bare $U$ is substantially smaller in GaMo$_4$S$_8$, the screening totally inverts this tendency. This can be again understood by considering the electronic structure of GaV$_4$S$_8$ and GaMo$_4$S$_8$ (Fig.~\ref{fig.LDA}): in GaMo$_4$S$_8$, the energy gaps separating $t_{2}$ and other bands are larger. Therefore, the screening should be weaker and the parameter $U$ itself -- larger.

\section{\label{sec:smodel} Spin model}
\par Details of the SE theory for the exchange interactions and the electric polarization can be found in Refs.~\cite{PRB2019,ROno}. In this theory, the kinetic energy in the leading order of $\hat{t}_{ij}/U$ is mapped onto the spin model $\mathcal{H}^{\mathrm{S}} = \sum_{\langle ij \rangle} \boldsymbol{e}_{i} \tensor{\mathscr{J}}_{ij} \boldsymbol{e}_{j}$, which can be further rearranged in terms of the isotropic exchange constants $J_{ij}$, antisymmetric DM vectors $\boldsymbol{D}_{ij}$, and the traceless symmetric anisotropic tensors $\tensor{\Gamma}_{ij}$ as~\cite{NJP2009,PRB2019}
\begin{equation}
\mathcal{H}^{\mathrm{S}} = \sum\limits_{\langle i j\rangle} \left( - J_{ij}\boldsymbol{e}_{i}\boldsymbol{e}_{j} + \boldsymbol{D}_{ij} \boldsymbol{e}_{i}\times\boldsymbol{e}_{j} + \boldsymbol{e}_{i} \tensor{\Gamma}_{ij} \boldsymbol{e}_{j} \right).
\label{eq:spinmodel}
\end{equation}
\noindent It is important to note that under the parity inversion, $J_{ij}$, $\boldsymbol{D}_{ij}$, and $\tensor{\Gamma}_{ij}$ behave as the (true) scalar, pseudovector, and tensor, respectively. Furthermore, we would like to stress that for the spin $1/2$ there should be no single-site contributions, neither to the exchange energy nor to the magnetic dependence of the electric polarization~\cite{ROno}. The DM interactions for the in-plane ($j$$=$$1$-$6$) and out-of-plane ($j$$=$$1'$-$6'$) bonds can be written as~\cite{PRB2019}
\noindent
\begin{equation}
\boldsymbol{D}_{0j} = d_{\parallel} \left( \sin \frac{\pi j}{3}, \cos \frac{\pi j}{3}, (-1)^{j} \delta \right)
\label{eq:DMin}
\end{equation}
\noindent and
\begin{equation}
\boldsymbol{D}_{0j} = d_{\perp} ( \cos \frac{\pi j}{3}, \sin \frac{\pi j}{3}, 0 ),
\label{eq:DMout}
\end{equation}
\noindent respectively, where the positions $j$ of the $M_4$ clusters are explained in Fig.~\ref{fig.notations}. The tensor $\tensor{\Gamma}_{ij}$ is given by
\noindent
\begin{widetext}
\begin{equation}
\tensor{\Gamma}_{0j} =
\left(
\begin{array}{ccc}
-\frac{1}{3} \Gamma + \Delta \Gamma \cos \frac{2\pi j}{3} & \pm \Delta \Gamma \sin \frac{2\pi j}{3}                   & \pm \Delta \Gamma' \sin \frac{2\pi j}{3} \\
\pm \Delta \Gamma \sin \frac{2\pi j}{3}                   & -\frac{1}{3} \Gamma - \Delta \Gamma \cos \frac{2\pi j}{3} & \Delta \Gamma' \cos \frac{2\pi j}{3} \\
\pm \Delta \Gamma' \sin \frac{2\pi j}{3}                  & \Delta \Gamma' \cos \frac{2\pi j}{3}                      & \frac{2}{3} \Gamma
\end{array}
\right),
\label{eq:Gamma}
\end{equation}
\end{widetext}
where the $+$ ($-$) signs stand for the in-plane (out-of-plane) bonds. The obtained parameters of the SE interactions are listed in Tables~\ref{tab:SEin} and~\ref{tab:SEout}. One can clearly see that in GaV$_4$S$_8$: (i) the isotropic exchange in and between the planes is clearly the strongest; (ii) the DM interactions are considerably weaker and can be viewed as a perturbation leading to the spin-spiral or skyrmion phase; and (iii) the symmetric anisotropic interaction is even weaker and can be neglected~\cite{PRB2019}.

\par Nevertheless, in GaMo$_4$S$_8$, the situation is fundamentally different. First, the isotropic exchange interactions are somewhat weaker than in GaV$_4$S$_8$. This can be understood as follows: (i) The transfer integrals $s^{3}$ and $u^{3}$, which contribute to the ferromagnetic (FM) and antiferromagnetic (AFM) paths connecting $a_{1}$ and $e$ orbitals~\cite{KugelKhomskii}, are comparable in GaV$_4$S$_8$ and GaMo$_4$S$_8$ (see Tables~\ref{tab:tin} and \ref{tab:tout}); (ii) On the other hand, the $J/U$ ratio, which controls the strength of the FM contributions to the exchange coupling~\cite{KugelKhomskii}, is smaller in GaMo$_4$S$_8$; (iii) Furthermore, the transfer integral $\mathit{t}_{\parallel}^{1}$, which contributes solely to the AFM coupling, is clearly lager in GaMo$_4$S$_8$. This effect is partly counterbalanced by larger $U$ value in the denominator of SE interactions, which is also larger in GaMo$_4$S$_8$. Altogether, this yields smaller $J_{\parallel}$ and $J_{\perp}$ in the case of GaMo$_4$S$_8$. Second, the DM interactions and the symmetric anisotropic interactions between the planes are of the same order of magnitude as $J_{\parallel}$ and $J_{\perp}$, as expected for materials with large SO coupling. Thus, in GaMo$_4$S$_8$ all interactions are comparable, which has a profound effect on the magnetic properties.
\noindent
\begin{table}[h!]
\caption{Parameters of superexchange interactions for the in-plane bonds (in meV). The corresponding vectors of Dzyaloshinskii-Moriya interactions and tensors of exchange anisotropy are given by Eq.~(\ref{eq:DMin}) and (\ref{eq:Gamma}), respectively.}
\label{tab:SEin}
\begin{ruledtabular}
\begin{tabular}{lrrrrrr}
              & $J_{\parallel}$ & $d_{\parallel}$ & $\delta$ & $\Gamma_{\parallel}$ & $\Delta  \Gamma_{\parallel}$ & $\Delta \Gamma_{\parallel}'$ \\
\hline
GaV$_4$S$_8$  & $0.180$         &   $0.073$       &  $0.137$ & $-0.007$             &   $-0.022$                   &    $0.003$       \\
GaMo$_4$S$_8$ & $0.110$         &   $0.179$       & $-0.399$ &  $0.004$             &   $-0.098$                   &   $-0.054$
\end{tabular}
\end{ruledtabular}
\end{table}
\noindent
\begin{table}[h!]
\caption{Parameters of superexchange interactions for the out-of-plane bonds (in meV). The corresponding vectors of Dzyaloshinskii-Moriya interactions and tensors of exchange anisotropy are given by Eq.~(\ref{eq:DMout}) and (\ref{eq:Gamma}), respectively.}
\label{tab:SEout}
\begin{ruledtabular}
\begin{tabular}{lrrrrr}
              & $J_{\perp}$ & $d_{\perp}$ & $\Gamma_{\perp}$ & $\Delta  \Gamma_{\perp}$ & $\Delta \Gamma_{\perp}'$ \\
\hline
GaV$_4$S$_8$  & $0.217$     & $0.057$     &         $-0.022$ &   $0.029$                &         $0$              \\
GaMo$_4$S$_8$ & $0.157$     & $0.136$     &         $-0.174$ &   $0.203$                &     $0.009$
\end{tabular}
\end{ruledtabular}
\end{table}

\par Very recently, the magnetic interactions in GaMo$_4$S$_8$ have been theoretically studied by mapping the total energies obtained in the generalized gradient approximation plus $U$ (GGA$+$$U$) onto the spin model~\cite{GaMo4S8Picozzi}. In principle, GGA$+$$U$ is the all-electron method and, in addition to the target $t_{2}$ bands, treats other valence states on an equal footing. On the other hand, the on-site Coulomb and exchange interactions in the GGA$+$$U$ method were treated in the basis of atomic Mo $4d$ orbitals, which we believe is a crude approximation and our choice of molecular Wannier basis for these purposes is more physical. Nevertheless, we note a qualitative agreement between our results and the ones of Ref.~\cite{GaMo4S8Picozzi}: at least in both studies $J_{\parallel} < J_{\perp}$, while $d_{\parallel} > d_{\perp}$ (note also the flip of the direction of the $z$ axis in Ref.~\cite{GaMo4S8Picozzi} in comparison with our choice of the coordinate frame, which should change the signs of $d_{\parallel}$ and $d_{\perp}$). However, the absolute values of the parameters of isotropic and DM interactions reported in Ref.~\cite{GaMo4S8Picozzi} are at least three times larger than ours. The direct comparison of the exchange anisotropy is ambiguous because the authors of Ref.~\cite{GaMo4S8Picozzi} have included in their analysis the unphysical single-site anisotropy term, which should vanish for the spin $1/2$.

\par In order to appreciate the importance of anisotropic interactions, it is instructive to estimate the Curie temperature, $T_{\rm C}$, using Tyablikov's RPA technique~\cite{tyab}. Then, considering only isotropic exchange interactions, we find $T_{\rm C} =$ $22$ and $10$~K for GaV$_4$S$_8$ and GaMo$_4$S$_8$, respectively. Naturally, since the nearest-neighbor interactions $J_{\parallel}$ and $J_{\perp}$ are larger in GaV$_4$S$_8$, the obtained $T_{\rm C}$ is also larger. Nevertheless, the experimental data reveal exactly the opposite tendency for $T_{\rm C}$. The discrepancy can be resolved by considering the anisotropic exchange interactions. Let us start with a simple semi-quantitative analysis of their effect. Among anisotropic exchange interactions, $\Gamma_{\perp}<0$ plays a very important role for the uniaxial systems, as it opens the magnon gap of a classical origin, which further increases $T_{\rm C}$~\cite{tyab}. $\Gamma_{\perp}$ is clearly one of the strongest interactions in GaMo$_4$S$_8$. Although $\Delta  \Gamma_{\perp}$ is formally comparable with $\Gamma_{\perp}$, it is typically responsible for a much smaller in-plane gap generated by quantum fluctuations~\cite{Yildirim}. Thus, as the first approximation, one can neglect $\Delta  \Gamma$ and evaluate $T_{\rm C}$ by considering only the isotropic exchange $J$ and the uniaxial anisotropy $\Gamma$, again in the framework of Tyablikov's RPA~\cite{tyab}. Quite expectedly, $T_{\rm C}$ practically does not change in the case of GaV$_4$S$_8$, where $\tensor{\Gamma}$ is small. In GaMo$_4$S$_8$, however, $\Gamma_{\perp}$ has a profound effect on $T_{\rm C}$, which increases to $22$ K and becomes comparable with the experimental value of $19$ K. This simplified analysis is fully supported by straightforward MC calculations for the model (\ref{eq:spinmodel}), which yield $T_{\rm C} \sim 18$ K for GaMo$_4$S$_8$ (in comparison with $T_{\rm C} \sim 23$ K for GaV$_4$S$_8$), as explained in Appendix~\ref{sec:MC}. Thus, we believe that relatively high $T_{\rm C}$ in GaMo$_4$S$_8$ is not because the isotropic exchange interactions are larger, but rather because the uniaxial anisotropy is stronger.

\par Finally, we note that the exchange parameters and $T_{\rm C}$ are sensitive to approximations employed for the solution of the effective low-energy model (\ref{eq:elmodel}) and definitions of the spin model. For instance, the SE approximation seems to overestimate $T_{\rm C}$ in GaV$_4$S$_8$ by factor 2 in comparison with the experimental value. In Appendix~\ref{sec:HF} we will show that, to certain extent, this discrepancy can be resolved by going beyond the SE approximation.

\par The spin-dependent part of the electric polarization can be written as~\cite{PRB2019,ROno} $\boldsymbol{P} = \sum_{\langle ij \rangle} \boldsymbol{\epsilon}_{ji} \left( \boldsymbol{e}_{i}\tensor{\mathscr{P}}_{ij}\boldsymbol{e}_{j} \right)$ or
\noindent
\begin{equation}
\boldsymbol{P} = \sum\limits_{\langle i j\rangle} \boldsymbol{\epsilon}_{ji} \left( P_{ij}\boldsymbol{e}_{i}\boldsymbol{e}_{j} + \boldsymbol{\mathcal{P}}_{ij} \boldsymbol{e}_{i}\times\boldsymbol{e}_{j} + \boldsymbol{e}_{i} \tensor{\Pi}_{ij} \boldsymbol{e}_{j} \right),
\label{eq:spinpol}
\end{equation}
\noindent where $\boldsymbol{\epsilon}_{ji} = \boldsymbol{\tau}_{ji}/|\boldsymbol{\tau}_{ji}|$ is the unit vector in the direction of the bond $i$-$j$ ($\boldsymbol{\tau}_{ji}=\boldsymbol{R}_{j}-\boldsymbol{R}_{i}$ being the bond vector connecting two $M_4$ clusters~\cite{PRB2019}). This is an analogue of Eq.~(\ref{eq:spinmodel}) for the electric polarization, where $\boldsymbol{\epsilon}_{ji} P_{ij}$, $\boldsymbol{\epsilon}_{ji} \boldsymbol{\mathcal{P}}_{ij}$, and  $\boldsymbol{\epsilon}_{ji} \tensor{\Pi}_{ij}$ stand for isotropic, antisymmetric, and anisotropic symmetric contributions, respectively. Alternative derivations for the isotropic and antisymmetric terms can be found in Refs.~\cite{pol2} and~\cite{superpol}, respectively. Importantly, since $\boldsymbol{P}_{ij} \parallel \boldsymbol{\epsilon}_{ji}$, \emph{only the out-of-plane bonds can contribute to the polarization change along} $z$. Therefore, we have to consider only the contributions of the sites $j$$=$$1'$-$6'$ (see Fig.~\ref{fig.notations}). Because of the additional prefactor $\boldsymbol{\epsilon}_{j0}$, the parameters $P_{0j}$, $\boldsymbol{\mathcal{P}}_{0j}$, and $\tensor{\Pi}_{0j}$  behave as, respectively, pseudoscalar, vector, and pseudotensor. Therefore, they will have the same form as $J_{0j}$, $\boldsymbol{D}_{0j}$, and $\tensor{\Gamma}_{0j}$, but multiplied by the additional prefactor $(-1)^{j}$. Thus, we get
\noindent
\begin{equation}
P_{0j} = (-1)^{j} P_{\perp},
\label{eq:Piso}
\end{equation}
\noindent
\begin{equation}
\boldsymbol{\mathcal{P}}_{0j} = (-1)^{j} p_{\perp} ( \cos \frac{\pi j}{3}, \sin \frac{\pi j}{3}, 0 ),
\label{eq:Pantisym}
\end{equation}
\noindent and
\noindent
\begin{widetext}
\begin{equation}
\tensor{\Pi}_{0j} =
(-1)^{j} \left(
\begin{array}{ccc}
-\frac{1}{3} \Pi + \Delta \Pi \cos \frac{2\pi j}{3} & -\Delta \Pi \sin \frac{2\pi j}{3}                   & -\Delta \Pi' \sin \frac{2\pi j}{3} \\
-\Delta \Pi \sin \frac{2\pi j}{3}                   & -\frac{1}{3} \Pi - \Delta \Pi \cos \frac{2\pi j}{3} & \Delta \Pi' \cos \frac{2\pi j}{3} \\
-\Delta \Pi' \sin \frac{2\pi j}{3}                  & \Delta \Pi' \cos \frac{2\pi j}{3}                   & \frac{2}{3} \Pi
\end{array}
\right).
\label{eq:Psym}
\end{equation}
\end{widetext}

\par The obtained parameters are listed in Table~\ref{tab:SEPol}.
\noindent
\begin{table}[h!]
\caption{Parameters of spin-dependent electric polarization (in $\mu\mathrm{C/m}^{2}$).}
\label{tab:SEPol}
\begin{ruledtabular}
\begin{tabular}{lccccc}
              & $P_{\perp}$      & $p_{\perp}$     & $\Pi_{\perp}$ & $\Delta \Pi_{\perp}$ & $\Delta \Pi_{\perp}'$ \\
\hline
GaV$_4$S$_8$  & $-362$           & $\phantom{-}41$ &      $1$      & $\phantom{-}7$       & $\phantom{-}1$        \\
GaMo$_4$S$_8$ & $\phantom{-}342$ & $-40$           &      $3$      & $-20$                & $-4$
\end{tabular}
\end{ruledtabular}
\end{table}
\noindent The contribution of $\tensor{\Pi}_{ij}$ to the polarization change associated with the skyrmion order is small (being of the 3rd order in $\zeta_{SO}$, as the angle between neighboring spins formed by the DM interactions is of the first order in $\zeta_{SO}$)~\cite{PRB2019}. Nevertheless, $P_{ij}$ and $\boldsymbol{\mathcal{P}}_{0j}$ can produce quiet comparable contributions to the polarization change in the 2nd order of $\zeta_{SO}$: $P_{ij}$ does not depend on $\zeta_{SO}$, but the change of $\boldsymbol{e}_{i}\boldsymbol{e}_{j}$ is of the 2nd order in $\zeta_{SO}$, while $\boldsymbol{\mathcal{P}}_{0j}$ and the change of $\boldsymbol{e}_{i}\times\boldsymbol{e}_{j}$ are both of the 1st order in $\zeta_{SO}$.

\par An interesting aspect of the magnetic dependence of the electric polarization in GaV$_4$S$_8$ and GaMo$_4$S$_8$ is that the parameters $P_{\perp}$ and $p_{\perp}$ of isotropic and antisymmetric contributions in these two compounds are comparable in absolute values, but have opposite signs, meaning that for the same spin texture, the polarization in GaV$_4$S$_8$ and GaMo$_4$S$_8$ will change in the opposite directions. This behavior can be rationalized by considering the analytical expression for $P_{ij}$~\cite{PRB2019}:
\noindent
\begin{equation}
P_{ij} \approx  \frac{e |\boldsymbol{\tau}_{ji}|}{V} \frac{J}{(U+|\Delta|)^3} \mathcal{T}_{ij},
\label{eq:Panalytical}
\end{equation}
where $\mathcal{T}_{ij} = (t_{ij}^{21})^{2}+(t_{ij}^{31})^{2}-(t_{ij}^{12})^{2}-(t_{ij}^{13})^{2}$ is antisymmetric with respect to the permutation of the atomic sites: $\mathcal{T}_{ij} =-\mathcal{T}_{ji}$. Then, using the analytical expression (\ref{eq:tout}) for the out-of-plane transfer integrals around $0$, one can find that $\mathcal{T}_{ij} = (-1)^{j+1} 2 \mathit{s}_{\perp}^{3} \mathit{u}_{\perp}^{3}$, which naturally explains that the reason why $P_{\perp}$ has opposite signs in GaV$_4$S$_8$ and GaMo$_4$S$_8$ is related to the opposite directions of the polar rhombohedral distortion, which controls the sign of $\mathit{u}_{\perp}^{3}$. Then, the transfer integrals and the SO interaction are progressively larger in GaMo$_4$S$_8$, which should lead to larger $P_{\perp}$ and $p_{\perp}$. Nevertheless, this effect is compensated by larger $U$ and $| \Delta |$ and smaller $J$, which reduces the value of the polarization in GaMo$_4$S$_8$ in comparison with GaV$_4$S$_8$. Similar tendencies are expected for $p_{\perp}$, as was confirmed by numerical calculations~\cite{PRB2019}.

\section{\label{sec:Phase} Emergence of skyrmions and change of electric polarization}
\par In order to study the formation of skyrmionic states, we perform MC simulations for the model~(\ref{eq:spinmodel}) in an external magnetic field parallel to $z$, $-\mu_{\rm B}h \sum_{i} e_{i}^{z}$. All technical details are summarized in Appendix~\ref{sec:MC}.

\par
\begin{figure}[b]
\begin{center}
\includegraphics[width=0.48\textwidth]{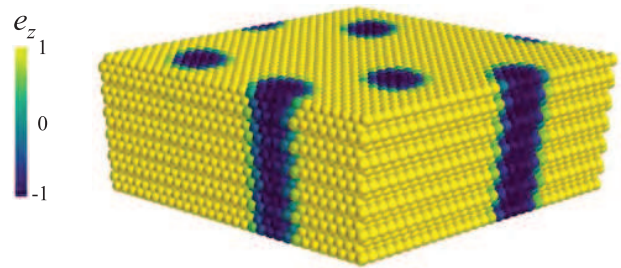}
\end{center}
\caption{Typical spin patterns obtained in Monte-Carlo simulation for the model (\ref{eq:spinmodel}) on the lattice $30\times30\times18$ ($\mu_{\rm B} h \sim 0.15 J_{\parallel}$ and $T=0.1J_{\parallel}$).}
\label{fig:tube}
\end{figure}

 The FM interlayer coupling $J_{\perp}$ tends to stack all two-dimensional spin patterns ferromagnetically along $z$. A typical ``tube structure'' obtained for the skyrmionic phase is illustrated in Fig.~\ref{fig:tube}.
\noindent

\par
\begin{figure}[b]
\begin{center}
\includegraphics[width=0.48\textwidth]{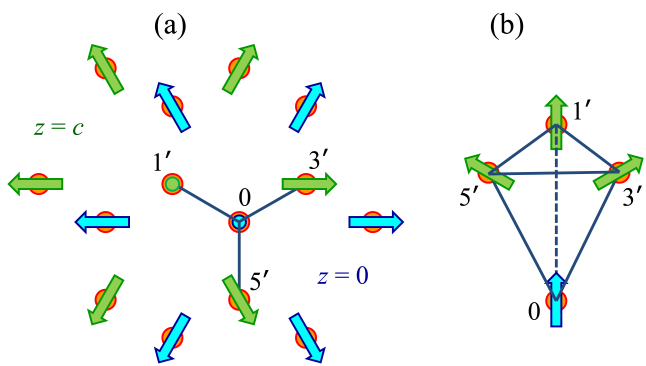}
\end{center}
\caption{Schematic illustration of interpenetrating skyrmionic textures in adjacent planes $z=0$ and $z=c$ with the notation of bonds formed by neighboring spins: (a) top view and (b) side view. Owing to the stacking misalignment, each spin couples with three neighboring spins in the next plane. Therefore, some of the neighboring spins between the planes will always be noncollinear.}
\label{fig:stacking}
\end{figure}

 The next important aspect is the stacking misalignment, which is inherent to the rhombohedral structure. Since only out-of-plane bonds contribute to the magnetic dependence of $P^z$, the change of the magnetic texture in the plane can affect this polarization only indirectly, via the redistribution of spins in adjacent planes. In this context, the ``stacking misalignment'' means that each next plane in the rhombohedral structure, besides the vertical shift along $z$, also experiences a horizontal displacement with respect to the original plane. Therefore, each spin couples with three neighboring spins in the next plane, meaning that some of these spins in the skyrmion tube will be noncollinear and the degree of this noncollinearity can be controlled by the magnetic field. According to our scenario, this is the main mechanism of the magnetic field dependence of $P^z$ in the lacunar spinel compounds~\cite{PRB2019}. The situation is schematically illustrated in Fig.~\ref{fig:stacking}: if the skyrmionic texture in the plane $z=c$ is obtained from the one in the plane $z=0$ by translating the spin $0$ to $1'$, the spins in the bonds $0$-$3'$ and $0$-$5'$ will still remain noncollinear.
\noindent

\par
\begin{figure}[t]
\begin{center}
\includegraphics[width=0.48\textwidth]{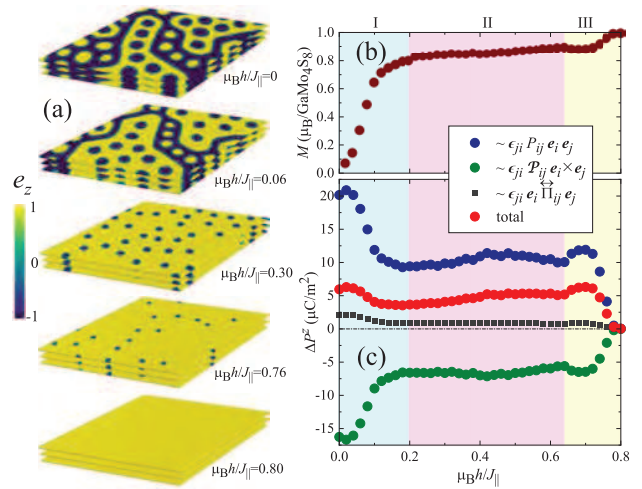}
\end{center}
\caption{(a) Spin patterns as obtained in Monte-Carlo calculations for the model (\ref{eq:spinmodel}) in an external magnetic field $h \parallel z$ at the temperature $T=0.1J_{\parallel}$. (b) Magnetization and (c) electric polarization (total and partial contributions) versus magnetic field. The meaning of the regions I, II, and III is explained in the text.}
\label{fig:phase}
\end{figure}

 Then, for each $h$ we obtain the distribution of spins and use it to evaluate the net magnetization and the change of electric polarization, $\Delta P^{z}$, relative to the FM state. The results are summarized in Fig.~\ref{fig:phase}.
\noindent

\par The phase diagram can be schematically divided in three regions. For small $h$ (region I), one can clearly see two FM domains corresponding to positive and negative directions of the magnetization along $z$. These domains are stabilized by DM interactions and their relative weight is controlled by the magnetic field. Furthermore, within each domain, one can clearly observe the skyrmions. In this case, the skyrmions are stabilized by strong uniaxial exchange anisotropy $\Gamma_{\perp}$, which plays the same role as the external field, but does not distinguish between positive and negative directions of the magnetization. This region corresponds to the rapid change of the electric polarization $P^{z}$, which mainly occurs at the AFM domain walls. Then, the system goes into the single domain region II. Nevertheless, the field $h$ still remains small compared to the exchange anisotropy $\Gamma_{\perp}$, which mainly controls the skyrmionic texture. As the result, the number and size of the skyrmions practically do not change, which is clearly manifested in the ``plateau'' of the magnetization and electric polarization versus $h$. The magnetic anisotropy energy due to $\Gamma_{\perp}$ can be evaluated as $\Delta E = 3 \Gamma_{\perp} \sim -$$0.5$ meV. Therefore, in order to produce a comparable effect, the magnetic field should be about $9$ T. In the region III, the magnetic field starts to prevail over the exchange anisotropy, and becomes the main factor controlling the size and the number of skyrmions. In this region, the magnetization strongly depends on $h$ and reaches the saturation in the FM state. The change of the magnetization is also accompanied by the rapid drop of the polarization.

\par Among three mechanisms of the polarization change -- the isotropic, antisymmetric, and symmetric anisotropic -- the latter is relatively weak, as was explained before. Then, there is a strong competition of isotropic and antisymmetric contributions to $\Delta P^{z}$, similar to GaV$_4$S$_8$~\cite{PRB2019}. These contributions enter with different signs and strongly cancel each other. Nevertheless, the isotropic term slightly dominates and controls the sign of total $\Delta P^{z}$ in both GaV$_4$S$_8$ and GaMo$_4$S$_8$. As the direction of the rhombohedral distortion changes, the sign of $\Delta P^{z}$ also changes when going from GaV$_4$S$_8$ ($\Delta P^{z} < 0$) to GaMo$_4$S$_8$ ($\Delta P^{z} > 0$). As was explained in Sec.~\ref{sec:smodel}, this is due to the behavior of parameters $P_{\perp}$ and $p_{\perp}$, which are odd functions of the rhombohedral distortion. It would be interesting to check this prediction experimentally.

\section{\label{sec:summary} Discussions and Summary}
\par Using first-principles electronic structure calculations, we have discussed the formation of skyrmions and the change of electric polarization, which is caused by these skyrmions in the lacunar spinel compounds GaV$_4$S$_8$ and GaMo$_4$S$_8$. For these purposes, we have constructed the effective electronic model for the molecular $t_{2}$ bands, which are located near the Fermi level and primarily responsible for the magnetism. The molecular character of the problem, where each magnetic lattice point is associated with the ($M_4$S$_4$)$^{5+}$ molecule, has a number of interesting consequences. Particularly, it is rather unusual, that the screened ``on-site'' Coulomb interaction $U$, characterising the repulsion of electrons within the ($M_4$S$_4$)$^{5+}$ molecules, is only of the order $0.7$ eV. For instance, in the atomic physics, such energy scale is characteristic for the Hund's exchange coupling $J$, while $U$ is expected to be substantially larger. Yet, in the molecular systems, the situation is different: $U$ is small and $J$ is even smaller (by an order of magnitude). Nevertheless, such ``small'' $U$ still remains to be the largest parameter in the problem, so that the transfer integrals, which are responsible for the dispersion of the $t_{2}$ bands can still be treated as a perturbation, in the spirit of the SE theory~\cite{Anderson}. We have successfully formulated such theory describing the behavior of exchange energy \emph{and} electric polarization in terms of relative orientation of spins in the bonds.

\par By using the spin model, obtained in the framework of the SE theory, we were able to rationalize the behavior of electric polarization in GaV$_4$S$_8$ and GaMo$_4$S$_8$. Particularly, although the Hund's coupling $J$ is small, it is the key parameter responsible for the magnetic dependence of $\boldsymbol{P}$, which is essentially the multiorbital effect being proportional to $J$ (and higher powers of $J$)~\cite{ROno}. Furthermore, in the SE theory, the electric polarization in each bond is always parallel to the direction of this bond. The division of magnetic solids into the bonds is an essential part of the SE concept: the energy is presented in terms of pairwise interactions occurring in the bonds~\cite{Anderson}. The same holds for the electric polarization. The new point here is that the bonds are polarized and can be views as electric dipoles. Moreover, the polarization of each such dipole depends on the relative orientation of spins in the bond.

\par Similar to the exchange energy, the magnetic dependence of the electric polarization in GaV$_4$S$_8$ and GaMo$_4$S$_8$ can be decomposed into isotropic, antisymmetric, and symmetric anisotropic parts. The latter is generally small, while the change of electric polarization induced by the skyrmion order results from the competition of isotropic and antisymmetric terms, which come with opposite signs. This is pretty much similar to the formation of the skyrmions themselves, resulting from the competition of isotropic and antisymmetric DM interactions. The basic difference, however, is that the competition of the exchange interactions takes place in the skyrmion plane, while for the polarization parallel to the $z$ axis, more important is the stacking of the skyrmion planes and the magnetic alignment in the bonds, which connect these planes.

\par Besides these similarities, the new aspect of GaMo$_4$S$_8$ is the strong uniaxial exchange anisotropy. We expect that this anisotropy is primarily responsible for higher $T_{\rm C}$ in the case GaMo$_4$S$_8$. Furthermore, it facilitates the formation of skyrmions, acting as a molecular field parallel to $z$, but making them relatively unsensitive to the external field in the large part of the phase diagram. Finally, we predict the reversal of the magnetic dependence of $\boldsymbol{P}$ in GaMo$_4$S$_8$, which is related to the reversal of the direction of rhombohedral distortion.

\appendix
\section{\label{sec:MC} Details of Monte Carlo simulations}
\par To study magnetic properties of GaMo$_{4}$S$_{8}$ at an external magnetic field, we performed classical MC simulations for the model (\ref{eq:spinmodel}) based on heat-bath method combined with overrelaxation and Metropolis algorithm~\cite{MC}. We used periodic hexagonal supercells with the $c$ axis parallel to $z = [111]$ (in the cubic setting) containing up to $N=30\times30\times18$, $N=36\times36\times9$, and $N=72\times72\times3$ sites. A single run contained $0.5\cdot10^{6}$ steps of equilibration and $2\cdot10^{6}$ steps of statistical averaging. For the initial relaxation, the system was gradually cooled down from higher temperatures. The Curie temperature is associated with the peak of specific heat at zero magnetic field:
\begin{equation}
\frac{C_{v}}{k_{B}} = \beta^{2}\frac{\langle E^{2} \rangle -\langle E \rangle^{2}}{V},
\end{equation}
\noindent where $\langle...\rangle$ stands for the thermal average, $E$ is the magnetic energy, $\beta=1/k_{B}T$, and $V$ is the volume of the supercell.

\par
\begin{figure}[b]
\begin{center}
\includegraphics[width=0.48\textwidth]{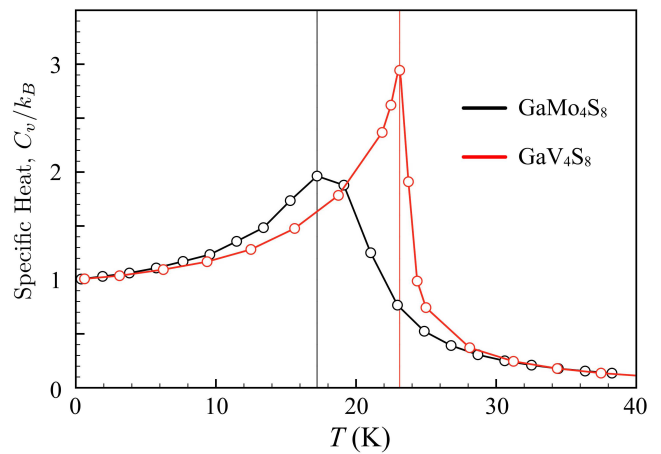}
\end{center}
\caption{Temperature dependence of the specific heat as obtained in the Monte Carlo calculations for the spin model (\ref{eq:spinmodel}) with the parameters derived in the superexchange approximation for GaV$_4$S$_8$ and GaMo$_4$S$_8$.}
\label{fig.MC}
\end{figure}

 The results of calculations for $C_{v}(T)$ are shown in Fig~\ref{fig.MC}.
\noindent
\noindent In order to take into account the quantum corrections, resulting from the replacement of $S^2$ by $S(S+1)$, the temperature in the classical Monte Carlo simulations was additionally scaled as $T \to (1+1/S)T$, similar to Tyablikov's RPA method~\cite{tyab}. Thus, we conclude that the theoretical $T_{\rm C}$, evaluated with parameters of SE interactions, is about $23$ and $18$ K for GaV$_{4}$S$_{8}$ GaMo$_{4}$S$_{8}$, respectively.

\section{\label{sec:HF} Alternative estimates of parameters of the spin model}
\par
\begin{figure}[b]
\begin{center}
\includegraphics[width=0.48\textwidth]{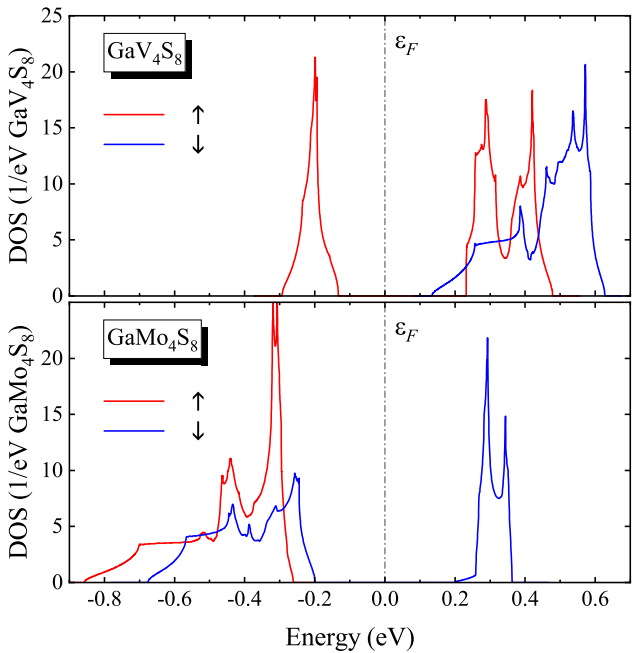}
\end{center}
\caption{Densities of states for the ferromagnetic state as obtained in the Hartree-Fock approximation for the model (\ref{eq:elmodel}).}
\label{fig.HFdos}
\end{figure}

 In this Appendix, we briefly discuss the results of the mean-field Hartree-Fock (HF) approximation for the solution of the model (\ref{eq:elmodel}) as an alternative to the SE theory for the exchange interactions and electric polarization. Fig.~\ref{fig.HFdos} shows the results of HF calculations without SO interaction for the densities of states in the FM phase.
\noindent
\noindent One can clearly see that the Coulomb repulsion, although being small, is sufficient for opening a band gap in both GaV$_4$S$_8$ and GaMo$_4$S$_8$. Nevertheless, the transfer integrals lead to the formation of bands whose width is at least comparable with the band gap, thus rising a question about applicability of the SE theory. Therefore, it is interesting to consider an alternative approach for the evaluation of exchange interactions and the electric polarization, which formally goes beyond the SE approximation.

\par Indeed, the parameters of interatomic exchange interactions can be very sensitive to the method and details of the electronic structure. The isotropic exchange interactions can be calculated using the theory of infinitesimal spin rotations near the FM state. The corresponding expressions in terms of the one-electron Green's function and intraatomic exchange field can be found in Refs.~\cite{JPCMreview,JHeisenberg}. The results are summarized in Table~\ref{tab:HF}.
\noindent
\begin{table*}[t]
\caption{Parameters of isotropic exchange ($J_{\parallel}$ and $J_{\perp}$), Dzyaloshisnkii-Moriya interactions ($d_{\parallel}$, $\delta$, and $d_{\perp}$), magnetocrystaline anisotropy energy ($\Delta E = E_{001} - E_{100}$), and anisotropy of electric polarization ($\Delta P = P_{001} - P_{100}$) as obtained in Hartree-Fock calculations.}
\label{tab:HF}
\begin{ruledtabular}
\begin{tabular}{lccccccc}
              & $J_{\parallel}$ (meV) & $J_{\perp}$ (meV) & $d_{\parallel}$ (meV) & $\delta$           & $d_{\perp}$ (meV)  & $\Delta E$ (meV) & $\Delta P$ ($\mu\mathrm{C/m}^{2}$) \\
\hline
GaV$_4$S$_8$  & $0.179$               & $0.330$           &      $0.124$          & $-0.076$           & $-0.029$           & $-0.041$          & $32$        \\
GaMo$_4$S$_8$ & $0.159$               & $0.027$           &      $0.107$          & $\phantom{-}0.081$ & $\phantom{-}0.230$ & $-0.306$          & $26$
\end{tabular}
\end{ruledtabular}
\end{table*}
\noindent In GaV$_4$S$_8$, $J_{\parallel}$ practically does not change in comparison with the SE calculations (see Table~\ref{tab:SEin}). However, $J_{\perp}$ increases by about 50\%. Nevertheless, this increase is accompanied by the appearance of six next nearest-neighbor AFM interactions between the planes, which were absent in the SE theory. These interactions are about $-0.152$ meV per bond, which overcompensate the increase of $J_{\perp}$. Furthermore, there are also small (about $-0.01$ meV) long-range AFM interactions in the plane. Altogether, it decreases stability of the FM states. The new $T_{\rm C}$, evaluated in the framework of Tyablikov's RPA but with the parameters extracted from the theory of infinitesimal spin rotations near the FM state, is about $10$ K~\cite{PRB2019}, which improves the agreement with the experiment ($T_{\rm C} \sim 13$ K~\cite{gavs2}). However, it should be understood that the theory of infinitesimal spin rotations probes mainly the stability of the ordered FM state. The present authors' opinion is that it is disputable whether the same model and parameters should describe the transition to the paramagnetic state, where the electronic structure is strongly affected by the spin disorder~\cite{Terakura}. Big changes are also expected in GaMo$_4$S$_8$, where in comparison with the SE theory $J_{\parallel}$ increases by about 50\%, while $J_{\perp}$ drops sharply by an order of magnitude (but still remains ferromagnetic).

\par The DM interactions can be evaluated using similar technique, in the first order of the SO coupling. For these purposes, it is convenient to use the self-consistent linear response theory, which takes into account the response of electron-electron interactions (\ref{eq:hublow}) onto the SO coupling in the HF approximation~\cite{solresp}. These calculations yield somewhat larger (smaller) value of $d_{\parallel}$ for GaV$_4$S$_8$ (GaMo$_4$S$_8$). Therefore, considering only the ratio $J_{\parallel}/d_{\parallel}$, it should lead to the some decrease (increase) of the skyrmion radii in GaV$_4$S$_8$ (GaMo$_4$S$_8$).

\par The uniaxial exchange anisotropy can be evaluated from the total energy difference $\Delta E = E_{001} - E_{100}$ between the out-of-plane and in-plane configurations of spins. Taking into account the definition (\ref{eq:spinmodel}) and the coordination numbers, one can find that $\Delta E = 3(\Gamma_{\parallel}+ \Gamma_{\perp})$. Then, assuming $| \Gamma_{\parallel} | \ll | \Gamma_{\perp} |$ (see Table~\ref{tab:SEin}), $\Gamma_{\perp}$ for GaV$_4$S$_8$ and GaMo$_4$S$_8$ can be estimated as $-0.014$ and $-0.102$ meV, respectively. These parameters are somewhat weaker than in the SE theory (see Table~\ref{tab:SEout}). Nevertheless, one can still conclude that $\Gamma_{\perp}$ is small and does not play any sizable role in GaV$_4$S$_8$, but expected to be important in GaMo$_4$S$_8$, where it becomes comparable with the parameters of isotropic and DM interactions, thus supporting our main conclusion obtained in the SE theory.

\par The only parameter, which can be easily derived by mapping the polarizations obtained in the HF calculations onto the spin model (\ref{eq:spinpol}) is $\Pi_{\perp}$, which is related with the calculated quantity $\Delta P = P_{001} - P_{100}$ as $\Delta P = 3 \epsilon^{z}_{01'} \Pi_{\perp}$, where $\epsilon^{z}_{01'}$ is $z$ component of the unit-vector $\boldsymbol{\epsilon}^{z}_{01'}$ ($\epsilon^{z}_{01'} = 0.819$ and $0.813$ for GaV$_4$S$_8$ and GaMo$_4$S$_8$, respectively). Thus, using results of Table~\ref{tab:HF}, $\Pi_{\perp}$ can be estimated as $13$ and $11$ $\mu\mathrm{C/m}^{2}$ for GaV$_4$S$_8$ and GaMo$_4$S$_8$, respectively: i.e., somewhat larger than in the SE theory, but still smaller in comparison with other parameters responsible for isotropic and antisymmetric contributions (see Table~\ref{tab:SEPol}).

\par In principle, other parameters of electric polarization, $P_{\perp}$ and $p_{\perp}$, can be also estimated by considering more complicated noncollinear magnetic textures and mapping results of the HF calculations onto the spin model (\ref{eq:spinpol}). Unfortunately, there is no analog of the theory of infinitesimal spin rotations for the electric polarization. Such extension would be certainly very interesting and helpful for the analysis of magnetoelectric coupling in various compounds.

\end{document}